\documentclass[conference]{IEEEtran}
\IEEEoverridecommandlockouts
\usepackage{color} 
\usepackage{bm}
\usepackage{multirow}
\usepackage{array} 
\usepackage{booktabs} 
\usepackage{cite}
\usepackage{amsmath,amssymb,amsfonts}
\usepackage{algorithmic}
\usepackage{graphicx}
\usepackage{textcomp}
\usepackage{xcolor}
\usepackage{stfloats}
\usepackage{float}
\usepackage{soul}
\usepackage{subcaption }
\usepackage{empheq} 
\def\BibTeX{{\rm B\kern-.05em{\sc i\kern-.025em b}\kern-.08em
    T\kern-.1667em\lower.7ex\hbox{E}\kern-.125emX}}
\colorlet{myhlcolor}{yellow!20}

\sethlcolor{myhlcolor}

\begin{document}
\title{Relevance Matters: A Multi-Task and Multi-Stage Large Language Model Approach for E-commerce Query Rewriting}

\author{Aijun Dai$^*$, Jixiang Zhang$^{\dag}$, Haiqing Hu$^{*}$, Guoyu Tang$^{*}$, Lin Liu$^{*}$, Ziguang Cheng$^{*}$\\
        \textit{$^*$JD.com, Beijing, China}\\
        \textit{$^{\dag}$Tsinghua University}}
\maketitle

\IEEEpubid{\makebox[\columnwidth]{
    \parbox{\columnwidth}{
        \vspace{60pt} 
        \scriptsize \copyright~2026 IEEE. Personal use of this material is permitted. Permission from IEEE must be obtained for all other uses, in any current or future media, including reprinting/republishing this material for advertising or promotional purposes, creating new collective works, for resale or redistribution to servers or lists, or reuse of any copyrighted component of this work in other works.
    }
} \hspace{\columnsep}\makebox[\columnwidth]{ }}

\begin{abstract}
For e-commerce search, user experience is measured by users' behavioral responses to returned products, like click-through rate and conversion rate, as well as the relevance between returned products and search queries. Consequently, relevance and user conversion constitute the two primary objectives in query rewriting, a strategy to bridge the lexical gap between user expressions and product descriptions. This research proposes a multi-task and multi-stage query rewriting framework grounded in large language models (LLMs). Critically, in contrast to previous works that primarily emphasized rewritten query generation, we inject the relevance task into query rewriting. Specifically, leveraging a pretrained model on user data and product information from JD.com, the approach initiates with multi-task supervised fine-tuning (SFT) comprising of the rewritten query generation task and the relevance tagging task between queries and rewrites. Subsequently, we employ Group Relative Policy Optimization (GRPO) for the model's objective alignment oriented toward enhancing the relevance and stimulating user conversions. Through offline evaluation and online A/B test, our framework illustrates substantial improvements in the effectiveness of e-commerce query rewriting, resulting in elevating the search results' relevance and boosting the number of purchases made per user (UCVR). Since August 2025, our approach has been implemented on JD.com, one of China's leading online shopping platforms.
\end{abstract}


\section{Introduction}

In the field of information retrieval, query rewriting is a highly effective approach to bridge the semantic gap between user expressions and search contents, helping users access information that meets their expectations. Consistently, in e-commerce search, query rewriting retrieves products by minimizing lexical discrepancies between user queries and product descriptions, thus better addressing user shopping needs and enhancing overall user experience. For e-commerce search, user experience is primarily evaluated by user interactions with search results and the relevance of retrieved items to the original queries. As a core component of e-commerce search systems, query rewriting is required to concurrently optimize two key objectives: user performance and perceived relevance.

In query rewriting, there are two main paradigms: discriminative methods and generative methods. Discriminative methods rephrase queries by retrieving similar terms, typically using sparse or dense retrieval techniques such as BM25 and Hierarchical Navigable Small World Graphs (HNSW) \cite{malkov2018efficientrobustapproximatenearest}. However, their reliance on exact lexical matching leads to semantic fragility and prevents them from recognizing semantic equivalence between different expressions.  In contrast, generative methods \cite{Qiu_2021, 10.1145/3219819.3219850}  use language models to directly rewrite queries into different forms. Both discriminative and generative methods overlook the importance of relevance in query rewriting, focusing exclusively on the lexical diversity of rewrites and user conversion rates. In the search retrieval funnel, query rewriting competes with alternative recall methods for a shared retrieval budget. Unrestrainedly increasing the textual diversity of rewrites can monopolize this retrieval budget, limiting the retrieval of relevant products from other pathways and significantly degrading the overall recall quality.

Recently, with the rapid advancement of large language models, an increasing number of studies have proposed LLM-based approaches to query rewriting. Most of these works \cite{dai2024enhancing, peng2024large} employ composite reward functions that incorporate relevance-oriented components. While relevance is included in the RL reward, the presence of competing signals within composite formulations tends to dilute its impact, constraining the LLMs’ ability to optimize relevance in a fine-grained manner during query rewriting. To address this limitation, our approach augments the fused RL reward with joint optimization of relevance tag prediction and query rewriting, thereby prompting deterministic relevance alignment.

\begin{figure*}[ht]
    \centering
    \includegraphics[width=0.9\textwidth]{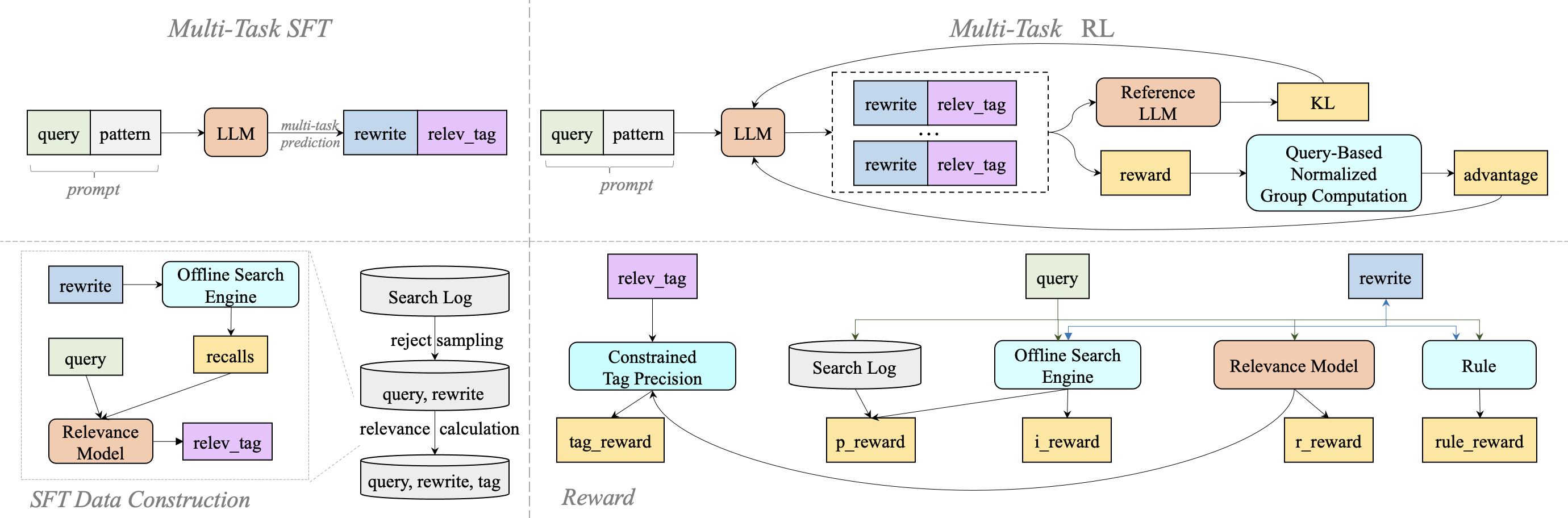}
    \caption{Multi-Task SFT and RL Architecture. (1)In SFT, LLM generates rewrites with the prompt as inputs and relev\_tag is constructed by Offline Search Engine and Relevance Model. (2)In RL, LLM is optimized by Query-Based Normalized GRPO. The integrated reward consists of tag reward, productive reward, increment reward, relevance reward and rule reward.}
    \label{fig:structure}
\end{figure*}

In this study, we propose an LLM-based multi-task supervised finetuning (SFT) and multi-task reinforcement learning (RL) framework shown in Fig\ref{fig:structure}, in which relevance prediction is integrated into query rewriting. Specifically, while generating rewrites, the post-pretrained LLM also produces corresponding relevance tags. For SFT training data construction, we extract query-rewrite pairs from online logs using rejection sampling. Then relevance scores are computed by the JD relevance model, measuring the relevance between original queries and products retrieved by rewrites; scores above a predefined threshold are assigned a relevance tag of 1, otherwise 0. During the SFT stage, the model jointly performs next-token prediction for both query rewrites and relevance tags, employing task-specific losses: cross-entropy for rewrite generation and binary cross-entropy for relevance tag prediction. In the RL stage, besides integrating rewards as in \cite{dai2024enhancing}, we introduce an additional reward for relevance tag prediction. The reward for rewrite generation is a composite of relevance, productivity, increment, and rule-based components. For relevance tag prediction, the reward is derived from the relevance scores between queries and products retrieved by the rewrites.

In summary, the primary contributions of this paper include:

\begin{itemize}

\item We propose a multi-task and multi-stage approach, integrating the relevance tag prediction into query rewriting. The approach significantly enhances the open-source LLMs' capabilities in e-commerce query rewriting, particularly relevance alignment between retrieved products and user intents.

\item We have introduced a reinforcement learning strategy that optimizes a composite reward function for query rewriting, achieving superior alignment of LLMs in e-commerce search compared to existing methods.

\item The efficacy of our framework has been corroborated through comprehensive offline evaluations and online A/B testing, confirming its practical value and robustness.

\end{itemize}

\section{RELATED WORK}
\subsection{Query Rewriting}

Query rewriting plays a pivotal role in e-commerce search technology to impact both the users' shopping experience and the revenue of e-commerce platforms. Broadly, query rewriting approaches can be categorized into two classes: discriminative methods and generative methods.

\textbf{Discriminative methods} frame query rewriting as a retrieval task, aiming to expand the semantics of the original query by selecting appropriate terms from a candidate set. Pseudo-relevance feedback approaches \cite{xu2017quary,10.1007/s10791-022-09405-y,naseri2022ceqe,zheng2020bert,khader2023learning,bassani2023personalized} typically enhance queries by identifying expansion terms from the top-ranked documents retrieved in an initial search, but may introduce noise and struggle to maintain relevance. To mitigate these issues, researchers have explored leveraging well-constructed thesaurus \cite{bhogal2007review,mandal2019query} or mining search logs \cite{antonellis2008simrank++,cui2002probabilistic,kunpeng2009new,li2022query,manchanda2019intent} to generate candidate rewrites. However, thesaurus-based methods depend heavily on coverage and quality. The log-based approaches are biased toward popular queries and lack sufficient data for long-tail queries.

\textbf{Generative methods} formulate query rewriting as a generative process by transformer-based models. Zero-grounding and non-interactive approaches \cite{wang2023query2doc,jagerman2023query,gao2023precise,liu2025exp4fuse} primarily rely on LLMs' prior knowledge and generate rewrites in a single pass, lacking domain-specific adaptation and resulting in limited retrieval effectiveness. Grounding-only and non-interactive methods \cite{jia2023mill,chen2024analyze,chuang2023expand,wang2023generative,lei2024corpus,zhang2024exploring,li2025pseudo,jaenich2025fair} explicitly condition generation on corpus evidence or pseudo-relevant documents with constraints or feedback to control semantic drift. Grounding-aware interactive approaches \cite{feng2023synergistic,rashid2024progressive,lei2025thinkqe} iteratively refine rewrites through multiple retrieve–rewrite cycles, but increasing engineering complexity.

\subsection{Objective Alignment}
Reinforcement Learning from Human Feedback (RLHF) \cite{ouyang2022training} represents a pivotal technique for aligning LLMs with human preferences and has been widely adopted for tasks such as safety assurance and reasoning enhancement. Proximal Policy Optimization (PPO) \cite{schulman2017proximal} is commonly employed in RLHF frameworks due to its stability and effectiveness. Direct Preference Optimization (DPO) \cite{rafailov2023direct} streamlines this process by directly optimizing model parameters using pairwise human preference data. Recently, Group Relative Policy Optimization (GRPO)~\cite{shao2024deepseekmath} and its variants \cite{yu2025dapo,liu2025understanding,zhang2025right,zheng2025group} have attracted significant attention, demonstrating promising results in objective alignment tasks. For query rewriting, several approaches inject domain-specific knowledge through pretraining or supervised fine-tuning (SFT), and further align model objectives using reinforcement learning algorithms \cite{dai2024enhancing,peng2024large,zuo2025value,zhang2025leser}. 

\section{METHOD}
\label{sec:method}
\subsection{Framework Overview}
Query Rewriting aims to enhance the semantics of the original query to bridge the semantic gap between user expressions and product descriptions, while ensuring user experience in which relevance is paramount. We propose a two-stage query rewriting framework consisting of multi-task SFT and multi-task RL alignment driven by the offline search engine. 

\subsection{Multi-task SFT}
\label{subsec:multi-task SFT}
Previous works have sought to bridge lexical gap between original query and its rewrite through carefully designed query rewriting generation methods. However, these approaches do not guarantee the relevance between the original query and its rewrite. To address this challenge, we inject the relevance prediction task into query rewriting, thereby ensuring both the effectiveness and relevance of the generated rewrites.

\textbf{Multi-task SFT dataset:} we initially collect query-rewrite pairs from search logs. Specifically, $x$ is a query, and the set of rewrites is $Y = \{y^1, y^2, \cdots, y^n\}$. We sample millions of user queries along with their associated rewrites to construct our initial query rewriting dataset $\mathcal{D}$:
\begin{equation}
\mathcal{D} = \left\{ 
    \left( x_{i}, y_{i}^k \right) \Big| \;
    x_{i} \sim p({x}),\;
    y_{i}^k \sim \pi_{\theta_{\text{old}}}\left( y \mid x = x_{i} \right),
    k=1...n
\right\}
\end{equation}

where $p(x)$ denotes the query distribution in our search engine, and $\pi_{\theta_{\text{old}}}$  represents the previous-generation rewriting policy of JD parameterized by $\theta_{\text{old}}$.

It is important that effective user clicks on the retrieved products of a rewrite serve as a key metric for measuring the relevance and the effectiveness of rewrite $y$. Therefore, we apply a click filter to the dataset $\mathcal{D}$ through rejection sampling:
\begin{equation}
\mathcal{D}_r = \left\{
    \left( x_{i}, y_{i}^k \right)\Big|\;\;
    (x_{i}, y_{i}^k) \in \mathcal{D},\;
    \mathcal{R}^{y_{i}^k} \cap \mathcal{C}^{x_{i}} \neq \emptyset 
\right\}
\end{equation}

where $\mathcal{R}^{y_{i}^k}$ is the recall set of rewrite $y_{i}^k$ and $\mathcal{C}^{x_{i}}$ is the click set of query $x_{i}$.

To further enhance the relevance between the generated rewrite and the original query, we incorporate the relevance tag prediction into the query rewriting process. Specifically, a relevance tag is assigned to each rewrite based on its relevance score:
\begin{equation}
\mathcal{D}_{sft} = \left\{
    \left( x_{i}, y_{i}^k, t )\right)\Big|\;\;
    (x_{i}, y_{i}^k) \in \mathcal{D}_r\;
    \right\}
\end{equation}
\begin{equation}
t =
    \begin{cases}
    1, & \text{if }r_R(x_{i}, \mathcal{R}^{y_{i}^k}) > \tau^{relev} \\
    0, & \text{otherwise}
    \end{cases}
\end{equation}
Here, $r_R(x, \mathcal{R}^y)$ is defined in Equation \ref{eq:relev}, and $\tau^{relev}$ denotes the relevance threshold. Experimental results show that increasing $\tau^{relev}$ leads to higher relevance but lower recall. In this work, $\tau^{relev}$ is set to 0.2.
 The SFT training task evolves from a single-task rewrite generation to a multi-task setting that jointly performs rewrite generation and relevance tagging. 


\textbf{Supervised Fine Tuning:} 
In the context of our multi-task SFT process, which includes both rewrite generation and relevance tagging, the training objective involves jointly minimizing the negative log-likelihood of the rewrite sequence and the binary cross-entropy of relevance tag prediction:
\begin{equation}
\begin{aligned}
\mathcal{L}_{\mathrm{SFT}}(\theta) = & -\mathbb{E}_{(x, y, t) \sim \mathcal{D}_{sft}} \left[ \sum_{i=1}^{|y|} \log \pi_{\theta} \left( y_i \mid y_{0:i-1}, x \right) \right. \\
& \left. + \lambda \left( t \log p_{\theta}(1 \mid x, y) + (1-t) \log p_{\theta}(0 \mid x, y) \right) \right]
\end{aligned}
\end{equation}
where $(x, y, t)$ denotes the query, the rewrite, and the relevance tag respectively; $\pi_{\theta}(\cdot)$ represents our query rewriting model parameterized by $\theta$, $p_{\theta}(t\mid x, y)$ is the predicted probability of the relevance tag, and $\lambda$ is a hyperparameter that balances the two tasks. 

\subsection{Multi-task RL Alignment}
Although LLMs achieve promising performance through SFT, RL can further improve the LLMs' ability \cite{guo2025deepseek} \cite{chu2025sft}. To directly refine the LLM’s query rewriting capabilities in the e-commerce domain, we employ GRPO to align the LLM’s objective toward enhancing relevance and stimulating user conversion, leveraging the offline search engine for alignment.


\textbf{Reward Design:} 
A well-designed reward function is essential for effective reinforcement learning\cite{gao2019reward}. We propose a hybrid reward that integrates both search engine feedback and rule-based components to enhance multi-task objectives.

\textit{Rewrite Reward:}
A central challenge in context-grounded query rewriting is the absence of ground-truth labels to supervise the training process. To address this, we optimize query rewriting by leveraging feedback from the offline search engine to guide and reward high-quality rewrites. The core intuition is that a high-quality rewrite should retrieve a set of products that is not only relevant to the original query but also diverse in its retrieval coverage. This principle enables us to design a feedback-based reward for rewrites, comprising three components derived from the offline search engine: 1) the relevance reward $r_R$, which measures the relevance between the query and the rewrite, 2) the productive reward $r_P$, which assesses the effectiveness of the rewrite, and 3) the increment reward $r_I$, which quantifies the complementary value of the rewrite. $r_{Feedback}$ is a hybrid combination of these individual parts.
\begin{equation}
\label{eq:relev}
r_R(x, \mathcal{R}^y) = \prod_{j=1}^{M} relev(x, \mathcal{R}_j^y)
\end{equation}

\begin{equation}
r_P(\mathcal{C}^x, R^y) = \sum_{i=1}^{N} \eta(\mathcal{C}_i^x, \mathcal{R}^y) \delta(\mathcal{C}_i^x, \mathcal{R}^y)
\end{equation}

\begin{equation}
\eta(\mathcal{C}_i^x, \mathcal{R}^y) = 
\begin{cases} 
1, & \text{if } \mathcal{C}_i^x \in \mathcal{R}^y \\
0, & \text{else}
\end{cases}
\end{equation}

\begin{equation}
r_I(\mathcal{R}^x, \mathcal{R}^y) = \frac{|\mathcal{R}^y| - |\mathcal{R}^x \cap \mathcal{R}^y|}{|\mathcal{R}^x|}
\end{equation}  
\begin{equation}
r_{Feedback}(x, y) = r_R(x, \mathcal{R}^y) \times r_I(\mathcal{R}^x, \mathcal{R}^y) + r_P(\mathcal{C}^x, \mathcal{R}^y)
\end{equation}
where $x$ is the query and $y$ is the rewrite, and accordingly $\mathcal{C}^x$ denotes the click set of query, $\mathcal{R}^x$ is the recall set of query and $\mathcal{R}^y$ is the
recall set of rewrite. $relev(\cdot)$ is the relevance score from the online relevance model of JD search. $|\cdot|$ is the counting operation of sets. $\delta(\cdot)$ is the recall position weight, which is set to 1, 2, or 3 depending on whether the click’s recall position is within the top 100\%, 70\%, or 30\%, respectively.

Furthermore, to promote retrieval diversity, we introduce a character-level difference rate reward that quantifies the diversity of rewrites relative to the original query. Additionally, a rewrite length constraint reward is incorporated to discourage overly long outputs and simple term appending. These rule-based rewards help guide the model toward generating well-formed and executable rewrites: 
\begin{equation}
r_{Rule}(x, y) = \alpha \times \mathrm{diff}(x,y) \times \frac{len(x)}{len(y)}
\end{equation}
where $\alpha$ is a hyparameter controlling the balance between reward components. Thus the final rewrite reward for original query $x$ and it's rewrite $y$ is defined as:
\begin{equation}
r_{Rewrite}(x, y) = r_{Rule}(x, y)\times r_{Feedback}(x, y)
\end{equation}

\textit{Constrained Tagging Reward:} We enforce a strict validation mechanism by assigning the final reward of zero to any output that does not satisfy this structural requirement. Based on this constraint, if the output adheres to the required format, the constrained tagging reward is defined as follows: 
\begin{equation}
r_{{Tag}}(x, y, t) =
\begin{cases}
1, & \text{if } |t - \mathbb{I}(r_R(x, \mathcal{R}^y) > \tau^{\mathrm{relev}})| = 0 \\
0, & \text{otherwise}
\end{cases}
\end{equation}
where $\mathbb{I}(\cdot)$ denotes the indicator function, $r_R(x, \mathcal{R}^y)$ is defined in Equation \ref{eq:relev} and $\tau^{relev}$ represents the relevance threshold. 

The final fusion reward is a hybrid signal that combines both search engine feedback and rule-based components to reinforce the multi-task objective:
\begin{equation}
\label{re:final_reward}
r_{Fusion}(x, y, t) =r_{Rewrite}(x, y) \times r_{Tag}(x, y, t)
\end{equation}
\textbf{Group Relative Policy Optimization (GRPO):} Given the one-to-many nature of query rewriting, where multiple rewrites may be valid for a single query, we adopt GRPO to fine-tune the model using relative rewards across candidate rewrites. For each query $x$, the policy $\pi_\theta(\cdot)$ samples $N$ candidates, denoted as $\hat{\bm{Y}} = \{\hat{y}^1, \ldots, \hat{y}^N\}$, where $\hat{y}^i$ represents a rewrite  $y^i$  paired with its corresponding relevance tag. Notably, beam search sampling is employed in this phase rather than top-$k$ or top-$p$ sampling.  The reward $r^i$ for each candidate $\hat{y^i}$ is computed with respect to the original query $x$ as defined in Equation \ref{re:final_reward}. The group-relative advantage for candidate $\hat{y}^i$ is then calculated as follows:
\begin{equation}
    A^i = \frac{r^i - mean(r^1,\dots, r^N)}{std(r^1,\dots, r^N)}
\end{equation}

 Since each rewrite $\hat{y^i}$ is serialized into a single output sequence, the likelihood of generating the rewrite is simply the sequence likelihood under the model. The probability ratio is defined as:
 \begin{equation}
  \rho^{i} = \frac{\pi_\theta(\hat{y}^i|x)}{\pi_{\theta_{\text{old}}}(\hat{y}_i|x)}   
 \end{equation}
 where both numerator and denominator are computed as products of token-level probabilities over the entire sequence $\hat{y}^i$. Finally, the GRPO objective combines clipped policy optimization with a KL regularizer:


\begin{equation}
\begin{aligned}
\mathcal{L}_{\mathrm{GRPO}}(\theta) =\; & -\mathbb{E}\big[x \sim P(x),\, \{\hat{y}^i\}_{i=1}^N \sim \pi_{\mathrm{old}}(\hat{y}|x)\big] \\
& \frac{1}{N} \sum_{i=1}^{N} \Bigg( \min\left(\rho^i A^i,\; \mathrm{clip}\left( \rho^i,\, 1-\epsilon,\, 1+\epsilon \right) A^i \right) \\
& \quad - \beta \mathbb{D}_{\mathrm{KL}}(\pi_\theta \| \pi_{\mathrm{ref}}) \Bigg)
\end{aligned}
\end{equation}
where the KL divergence is estimated using an unbiased estimator:
\begin{equation}
\mathbb{D}_{\mathrm{KL}}(\pi_\theta \| \pi_{\mathrm{ref}}) =\; \frac{\pi_{\mathrm{ref}}(\hat{y}^i|x)}{\pi_\theta(\hat{y}^i|x)} - \log \frac{\pi_{\mathrm{ref}}(\hat{y}^i|x)}{\pi_\theta(\hat{y}^i|x)} - 1
\end{equation}

In this objective, $\epsilon$ specifies the clipping range for stabilizing updates, while $\beta$ controls the strength of the KL regularization.

\subsection{Online Serving}
In the context of deploying LLMs for online serving, it is crucial to consider both computational efficiency and latency. We adopt a near-line deployment strategy \cite{amatriain2015recommender} to meet the latency requirements of the JD search system while fully leveraging the capabilities of the LLM. For each new query, our query rewriting LLM generates a list of rewrites along with corresponding relevance tags using beam search sampling. After filtering based on relevance tags, the top three rewrites are selected as the final rewrites. Once the rewrites are generated, they are cached for up to 14 days to enable rapid online responses. 

\section{EXPERIMENTS}

\subsection{Datasets}
\textbf{Multi-task SFT Dataset:}
As detailed in Section \ref{subsec:multi-task SFT}, we obtain a multi-task SFT dataset consisting of approximately 5.6 million queries and rewrites, each annotated with its corresponding relevance tag and table \ref{tab:data} demonstrates it.

\textbf{Multi-task RL Dataset:}
The RL dataset, comprising 36k queries along with their corresponding clicked product sets, is constructed from the SFT dataset using the RL reward defined in our algorithm.

\textbf{Evaluation Datasets:}
For the offline test set, a manually annotated dataset \mbox{\ref{tab:data}} of 1k $<$query, rewrite$>$ pairs is constructed to serve as ground-truth for evaluating the accuracy of relevance tagging.
Since rewrites retrieve products via inverted index methods, we construct a recall evaluation dataset employing embedding-based retrieval strategies. The resulting evaluation dataset comprises 100k queries and their corresponding retrieved product pairs, where the products have been frequently clicked by users for the given queries. 

\begin{table}[tb]
\centering
\small 
\caption{Prompt Examples and Data Examples. For SFT, the blue parts are alternative. For RL, "The synonymous search term and its corresponding relevance tag for \{query\}" is the input for LLMs to generate rewrite and relevance tag.}
\label{tab:data}
\begin{tabular}{>{\centering\arraybackslash}p{1.3cm}|p{6.5cm}}
\toprule
\textbf{Stage} & \textbf{Prompt Sample / Data Sample} \\ \hline
\multirow{2}{*}{SFT} & \textit{The synonymous search term and its corresponding relevance tag for \{\textcolor{blue}{query}\} are \{\textcolor{blue}{rewrite}\}} $<|sep|>$ \textit{\{\textcolor{blue}{relev\_tag}\}}\\ \cline{2-2}
 & The synonymous search term and its corresponding relevance tag for Kappa school bag are Kappa backpack $<|sep|>$ 1. \\ \hline
\multirow{2}{*}{RL} & \textit{query, product titles, click product ids}\\ \cline{2-2}
 & perfume for men, \{Men’s 50ml Eau de Toilette Oriental Fragrance, Tea Eau de Toilette 50ml Woody Fragrance\}, $\{id_1, id_2\}$ \\ \hline
\multirow{2}{*}{Evaluation} & \textit{query, product title}\\ \cline{2-2}
& baby yogurt, SIMPLELOVE Father's Love Formula Children's Yogurt 0\% Sucrose Original Flavor 100g*6 Bags \\ \bottomrule
\end{tabular}
\end{table}

\subsection{Evaluation}
\textbf{Offline Metrics:} To comprehensively evaluate our query rewriting approach, we utilize several offline metrics:

\begin{enumerate}
    \item \textbf{Tagging Accuracy:} An accuracy-oriented metric that measures the alignment between the relevance tags generated by our LLM and the ground-truth labels assigned by experienced annotators.
    \item \textbf{Recall Rate:} This metric evaluates the extent to which rewrites generated by the LLM benefit the search engine, defined as:
    \begin{equation}
    \text{recall} = \frac{\sum_{x \in \mathcal{D}_{recall}} |\mathcal{R}^{\mathcal{Y}} \cap \mathcal{C}^x|}{|\mathcal{D}_{recall}|}    
    \end{equation}
    where $\mathcal{D}_{recall}$ denotes the recall evaluation dataset, $\mathcal{Y}$ is the set of rewrites for $x$ filtered by relevance tag, $\mathcal{R}^{\mathcal{Y}}$ is recall set of $\mathcal{Y}$  obtained via inverted index methods and $\mathcal{C}^x$ is the valid clicked set retrieved by either item-based or embedding-based retrieval. $|\cdot|$ is the counting operation of sets. 
    \item \textbf{Relevance Score:} A semantic alignment metric that measures the average relevance score between rewrites and queries. The relevance score for each $<$query, rewrite$>$ pair is calculated as described in Equation (6).
\end{enumerate}
\textbf{Online Metrics:} To evaluate the model’s online performance, we use three metrics: user conversion rate (UCVR), user click-through rate (UCTR) and Unique Visitor Value(UV value).

\subsection{Implementation Details}
Both the SFT and RL training processes were conducted using the open-source TRL framework \mbox{\cite{vonwerra2020trl}}. Below, we give detailed explanations of key training parameters, while all unmentioned parameters were kept at their default values.

\textbf{Multi-task Fine-tuning:} We fine-tuned the Qwen3-8B model using a cosine learning rate schedule with a peak learning rate of 2e-5 and a warmup ratio of 0.03. The process was optimized with AdamW ($\beta_1 = 0.9$, $\beta_2 = 0.999$) and a weight decay of 0.001. We used a batch size of 64 and applied the DeepSpeed ZeRO Stage 1 parallelism strategy with bfloat16 precision. The model was trained for 5 epochs on 8 Ascend 910B NPUs for a total of 25 hours.


\textbf{Post-training Alignment:} The alignment phase was performed with a linear learning rate schedule set to 1e-6. We configured the AdamW optimizer with coefficients $(\beta_1, \beta_2)$ of $(0.9, 0.999)$ and omitted weight decay to maintain training consistency. The KL coefficient $\beta$ in GRPO was set to 0.4. The training was executed on 4 NVIDIA H800 GPUs for a duration of 18 hours.


\subsection{Offline Experiments}

\subsubsection{Comparison of Different LLMs}
In this section, we present a comparative analysis of different LLMs as base models trained on the multi-task SFT dataset, as summarized in Table \ref{table:comparison}. Several open-source LLMs are considered, including  BaiChuan~\cite{yang2023baichuan}, Qwen2.5~\cite{team2024qwen2}, 
Hunyuan \cite{tencent2025hunyuan7b}, Qwen3~\cite{yang2025qwen3} and R1-distill Qwen3. Among these, R1-distilled Qwen3-8B exhibits superior overall performance compared to the other models. However, Qwen3-8B achieves the highest relevance score, which is the primary focus of this work. Consequently, we select Qwen3-8B as our base model and further perform e-commerce domain-specific pretraining on it prior to subsequent experiments.
\begin{table}[htbp]
\centering
\caption{Comparison of different LLMs trained on multi-task SFT dataset.}
\label{table:comparison}
\begin{tabular}{lccc}
\toprule
\textbf{Model} & \textbf{Acc(\%)} & \textbf{Recall(\%)} & \textbf{Relev(\%)}  \\
\midrule
Baichuan-7B & 79.9 & 62.69 & 73.03  \\
Qwen2.5-7B & 80.5 & 62.79 & 72.90  \\
Hunyuan-7B & 80.7 & 62.85 & 71.57  \\
\textbf{Qwen3-8B} & 80.8 & 62.86 & 73.21  \\
R1-Qwen3-8B & 81.6 & 63.18 & 71.79  \\
\bottomrule
\end{tabular}
\end{table}

\subsubsection{Multi-task SFT V.S. Single-task SFT}
Given that the training objective of multi-task SFT is to jointly generate both the rewrite and its relevance tag in a single step, we aim to compare this approach with a sing-task SFT setting where the model is trained to generate only the rewrite.

\begin{figure}[htbp]
    \centering
    \begin{minipage}{0.24\textwidth}
        \includegraphics[width=\linewidth]{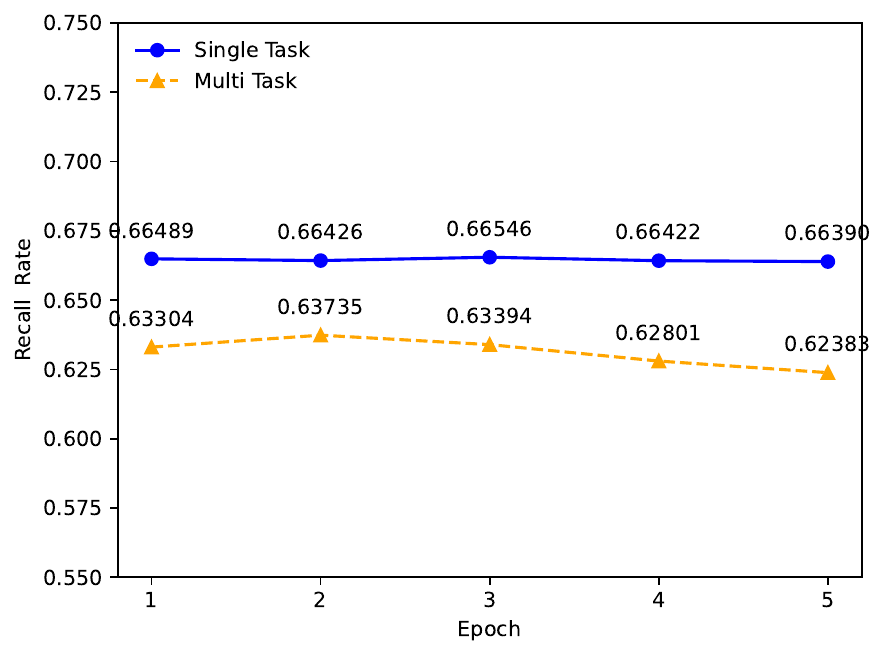}
    \end{minipage}
    \begin{minipage}{0.24\textwidth}
        \includegraphics[width=\linewidth]{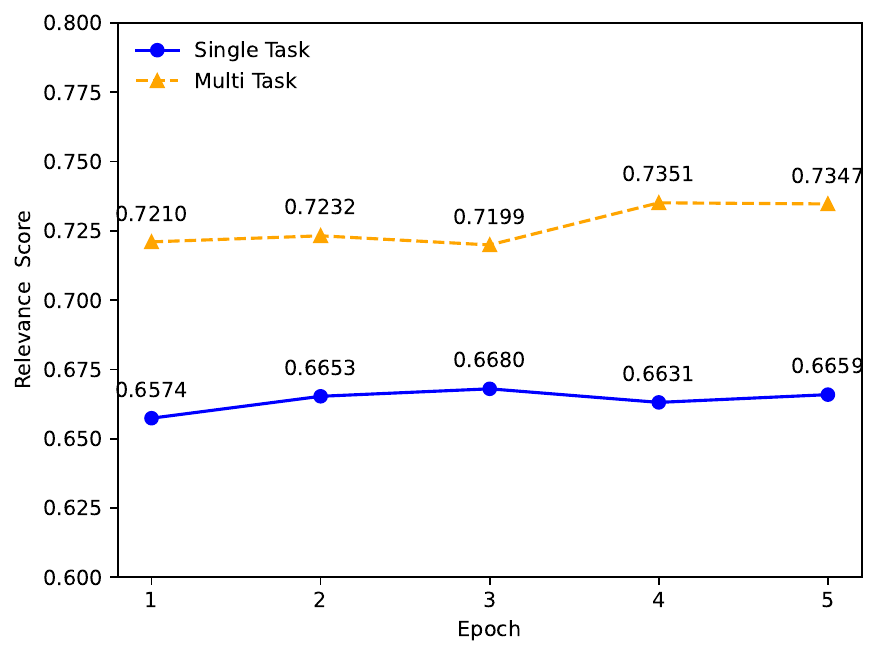}
    \end{minipage}
    \caption{Comparison of Recall Rate and Relevance Score}
    \label{fig:comparison}
\end{figure}

For single-task, we perform SFT using the $<$query, rewrite$>$ pairs obtained via rejection sampling as described in Section \ref{subsec:multi-task SFT}, with the model trained solely for rewrite generation. 
Figure \ref{fig:comparison} demonstrates that compared with model with single-task SFT, multi-task SFT achieve significantly better relevance but lower recall rates. 
These results highlight the effectiveness of multi-task SFT in enhancing rewrite relevance.


\subsubsection{Comprehensive Study of RL Algorithms}
Recent advances \mbox{\cite{peng2024large,zuo2025value,zhang2025leser}} show that reinforcement learning algorithms significantly contribute to performance improvements in query rewriting tasks. Building upon these insights, our comparative experiments primarily focuse on the reinforcement learning stage, where we added DPO and PPO baselines and performed detailed analyses on the individual and combined effects of different reward components within our proposed framework. We constructed 72k DPO training pairs by sampling 10 candidate rewrites per query(from RL dataset) via beam search with SFT model, and scoring them with the fusion reward. For each query, the top-scoring rewrite was chosen as the positive sample and two negatives were randomly selected from the five lowest-scoring candidates. For the PPO experiments, we directly used the fusion reward defined in our work as the
training reward.
\begin{table}[htbp]
\centering
\caption{Comprehensive Study of RL Algorithms}
\label{table:RL comparison}
\resizebox{\columnwidth}{!}{
        \begin{tabular}{clccc}
        \toprule
        & \textbf{Method} & \textbf{Acc(\%)} & \textbf{Recall(\%)} & \textbf{Relev(\%)} \\
        \midrule
        & pretrain+SFT  & 80.0 & 63.74 & 72.32 \\
        \midrule
        \multirow{9}{*}{\shortstack{Pretrain \\ + \\ SFT}} 
        & +DPO & 81.2 & 63.93 & 72.61 \\
        & +PPO & 80.9 & 65.89 & 70.03 \\
        \cmidrule{2-5}
        & +GRPO ($r_P$) & 77.2 & 64.04 & 74.17 \\
        & +GRPO ($r_I$) & 81.6 & 47.22 & 33.97 \\
        & +GRPO ($r_R$) & 77.4 & 34.07 & 84.15 \\
        & +GRPO ($r_{Feedback}$)& 77.8 & 62.39 & 68.98 \\
        & +GRPO ($r_{Rewrite}$)& 80.1 & 70.18 & 64.16 \\
        & +GRPO ($r_{{Feedback}\&{Tag}}$)& 81.9 & 59.36 & 73.29 \\
        & +GRPO ($r_{Fusion}$)& 81.5 & 67.94 & 69.65 \\
        \bottomrule
        \end{tabular}
}
\end{table}

As shown in Table \mbox{\ref{table:RL comparison}}, both DPO and PPO yield moderate improvements across overall metrics. For GRPO, each individual reward component effectively serves its intended purpose, but often at the expense of other performance metrics. The fused reward achieves the best overall performance. ($r_{Feedback\&Tag} = r_{Feedback} \times r_{Tag}$)

\subsubsection{Ablation Study of Beam Size In GRPO}
As discribed above, after multi-task SFT, our model achieves higher relevance score but lower recall rate. To enhance recall rate while maintaining relevance, we adopt GRPO for objective alignment. 
Notably, beam search sampling is employed in this phase rather than top-$k$ or top-$p$ sampling. Table \ref{table:ablation_study} presents the effect of varying beam sizes on model performance during the objective alignment stage. It can be observed that as the beam size increases, both recall rate and tagging accuracy consistently improve, while relevance score initially rises and subsequently declines. 

\begin{table}[htbp]
\centering
\caption{Ablation Study of Beam Size}
\label{table:ablation_study}
\begin{tabular}{lcccc}
\toprule
\textbf{Method} & \textbf{Beam Size} & \textbf{Acc(\%)} & \textbf{Recall(\%)} & \textbf{Relev(\%)} \\
\midrule
pretrain+SFT & / & 80.0 & 63.74 & 72.32 \\
\midrule
pretrain+SFT+GRPO & 3 & 62.9 & 66.76 & 65.86 \\
pretrain+SFT+GRPO & 5 & 68.7 & 67.30 & 66.05 \\
pretrain+SFT+GRPO & \textbf{10} & 78.5 & 67.94 & 69.65 \\
pretrain+SFT+GRPO & 15 & 78.9 & 68.97 & 67.79 \\
\bottomrule
\end{tabular}
\end{table}

\subsubsection{Result of Different Rewrite Number} The number of rewrites retained after filtering by relevance tags for each query also influences retrieval performance. It is straightforward that increasing the number of rewrite candidates leads to a notable improvement in recall rate. 

\begin{table}[htbp]
\centering
\caption{Result of Different Rewrite Number.}
\label{table:rewrite_number}
\begin{tabular}{ccc}
\toprule
\textbf{Rewrite Number} & \textbf{Recall(\%)} & \textbf{Relev(\%)} \\
\midrule
3 & 67.94 & 69.65 \\
4 & 69.44 & 69.65 \\
5 & 69.96 & 69.80 \\
\bottomrule
\end{tabular}
\end{table}
\subsection{Online Experiments}
To evaluate the online performance of our model, we deployed it in an online search engine and allocated 10\% of user traffic for a seven-day A/B testing period. The evaluation focused on three key metrics: Unique Visitor Value(UV value), user conversion rate (UCVR), and user click-through rate (UCTR). As shown in Table \ref{table:online_experiments}, our model outperformed the previous-generation rewriting model by 0.106\%, 0.181\%, and 0.082\% in terms of UV, UCVR, and UCTR, respectively. These results demonstrate that our model effectively rewrites queries and addresses potential semantic gaps in the semantic matching process. 
\begin{table}[htbp]
\centering
\caption{Result of Online Experiments.}
\label{table:online_experiments}
\begin{tabular}{cccc}
\toprule
\textbf{Online Traffic} & \textbf{UV value} & \textbf{uCVR} & \textbf{uCTR} \\
\midrule
All & +0.106\% & +0.181\% & +0.082\% \\
Top & +0.240\% & +0.170\% & +0.110\% \\
Torso & +0.430\% & +0.030\% & +0.020\% \\
Long-tail & +0.490\% & +0.290\% & +0.110\% \\
\bottomrule
\end{tabular}
\end{table}
\subsection{Case Study}
Table \ref{table:cases} reveals some case samples of our model’s results. Case1 and Case2 shows our multi-task framework enhance LLMs' comprehensive ability in e-commerce, where ``NB'' is the abbreviation for ``New Balance''. Case3 illustrates the effectiveness of introducing the relevance task into query rewriting, reducing the retrival of irrelevanr products.
\begin{table}[htbp]
\centering
\caption{Case Samples.}
\label{table:cases}
\begin{tabular}{cccc}
\toprule
\textbf{ } & \textbf{Query} & \textbf{Rewrite} & \textbf{Relev\_tag} \\
\midrule
1 & kitchen drain pipe & sink drain pipe & 1 \\
2 & NB down jacket & New Balance down jacket & 1 \\
3 & Xiaomi pad & tablet & 0 \\
\bottomrule
\end{tabular}
\end{table}

\section{CONCLUSION}
In this paper, we reveal a pivotal limitation in e-commerce query rewriting: the systematic underemphasis on relevance critically impacts retrieval quality. Consequently, we propose a multi-task and multi-stage LLM framework, integrating relevance tag prediction into query rewriting. In the SFT stage, LLM concurrently generates rewrites and corresponding relevance tags with task-specific losses. For objective alignment towards relevance and user conversion, we employ GRPO with query-based normalized group optimization for advantage calculation and a comprehensive reward with the relevance reward, the productive reward, the increment reward, the rule reward and the relevance tag reward. Offline and online experiments validate the efficacy of our multi-task and muli-stage framework.

\vspace{12pt}
\bibliographystyle{./IEEEtran}
\bibliography{./IEEEabrv,./sample_base}.

\end{document}